\documentstyle[aps,preprint,eqsecnum,epsf]{revtex}

\draft
\tightenlines

\begin{document}
\title{Baryon Magnetic Moments\\ in a QCD-based Quark Model
with loop corrections}
\author{Phuoc Ha\thanks{Electronic address: phuoc@theory1.physics.wisc.edu}
and  Loyal Durand\thanks{Electronic address:ldurand@theory2.physics.wisc.edu}}
\address{
        Physics Department, University of Wisconsin-Madison \\
             Madison, Wisconsin 53706, USA
        }
\date{\today}
\maketitle

\begin{abstract}
We study meson loop corrections to the baryon magnetic moments
starting from a QCD-based quark model derived earlier in a quenched
approximation to QCD. The model reproduces the standard quark model
with extra corrections for the binding of the quarks. The loop corrections
are necessary to remove the quenching. Our calculations use
heavy baryon perturbation theory with chiral baryon-meson couplings and
a form factor characterizing the structure of baryons as composite
particles. The form factor reflects soft wave function effects
with characteristic momenta $\approx 400$ MeV, well below the
usual chiral cutoff of $\approx 1$ GeV. The resulting model involves only
three parameters, the quark moments $\mu_u$ and $\mu_s$ and a parameter
$\lambda$ that sets the momentum scale in the wavefunctions. We find that
this approach substantially improves the agreement between the
theoretical and experimental values of the
octet baryon magnetic moments, with
an average difference between the theoretical and experimental moments
of 0.05$\mu_N$. An extension to the decuplet states using the same input
predicts a moment of 1.97 $\mu_N$ for the $\Omega^-$
hyperon, in excellent agreement with the measured moment of
$2.02\pm0.05\ \mu_N$.
\end{abstract}


\section{Introduction}

The simple, nonrelativistic quark model (QM) gives a qualitatively good
description of the baryon moments. Under the assumption that each baryon
is composed of three valence or constituent quarks in a state with all
internal orbital angular momenta equal to zero, the moments are given
by expectation values of the spin moment operators
\begin{equation}
\mbox{\boldmath$\mu$}^{QM}=\sum_{\rm q}\,\mu_{\rm q}
\mbox{\boldmath$\sigma$}_{\rm q},
\label{eq:mu}
\end{equation}
leading to the standard expressions
\begin{equation}
\mu_p^{QM}=\frac{1}{3}(4\mu_u-\mu_d),\,\ldots,\quad \mu_q=\frac{e_q}
{2m_q}\,, \label{eq:mu_quark}
\end{equation}
where the effective quark moments $\mu_u=-2\mu_d$ and $\mu_s$ can be
treated as free parameters in attempting to fit the data.
The predicted pattern of the signs for the moments agrees with observation.
A least-squares fit to seven measured octet moments (the transition moment
$\mu_{\Sigma^0\Lambda}$ is left to be predicted) gives a
root-mean-square deviation of theory from experiment of 0.12 $\mu_N$, about
10\% of the average magnitudes of the moments. Agreement at this level can be
regarded as an outstanding success of the quark model, but the deviations also
give a very sensitive test of baryon structure. There is presently no
completely successful first-principles theory of the moments.

In this paper, we approach the moment problem dynamically using a QCD-based
quark model \cite{Durand} with meson loop corrections.
The model has only three
parameters, namely the quark moments $\mu_u=-2\mu_d$ and $\mu_s$, and a
parameter $\lambda$ that characterizes the momentum scale of meson-baryon
interactions, with the particles regarded as composite. We find that this
approach reduces the average deviation of the calculated moments from
experiment
to $0.05 \mu_N$, a substantial improvement, and that the loop corrections
are small compared to the leading QM terms, suggesting reasonable convergence
for the loop expansion.

A different approach to the moment problem
using chiral perturbation theory (ChPT)
with loop corrections has been studied recently by a number of authors.
Excellent fits to the data can be obtained for either conventional
\cite{Jenetal,Luty,Daietal,Mei} or a modified \cite{Bos} chiral counting.
However, as pointed out elsewhere \cite{DandH}, the general fits
are essentially
independent of the dynamical input, and are best thought of as
giving parametrizations of the data. In particular, the introduction of
the counterterms needed to eliminate divergences in the loop integrals results
in the appearance of five new chiral couplings \cite{Jenetal,Mei,Bos}
at the one-loop level in addition
to the two tree-level \cite{Coleman} couplings.  The seven well-measured
octet moments can be fitted exactly using these seven parameters,
whether or not the calculated loop corrections are included \cite{DandH}.
The theory is only weakly predictive with respect to the remaining quantity,
the $\Sigma^0\Lambda$ transition moment, which is not known precisely.
Furthermore, the loop corrections in ChPT are nearly as large as the
tree-level terms, and convergence of the expansion is at best slow, when
the divergences are regularized using dimensional regularization.
This problem is not unexpected, since chiral symmetry is known to be
badly broken for the baryons. The remaining
predictions \cite{Jenetal,Luty,Daietal} are sum rules that can be motivated
by large $N_c$ expansions or expansions in the symmetry breaking mass
parameter $m_s$ in ChPT, but do not depend
explicitly on the dynamics.

Our intent here is to present a dynamical approach to the baryon moments
in which we emphasize the composite nature of the baryons and use a
description based on the QM rather than the chiral picture.
We begin in Sec.\ II by reviewing briefly the derivation of the QM,
with corrections, from QCD using a Wilson-loop approach \cite{Durand}.
The derivation involves the suppression of internal quark loops
(``quenching''), an approximation that is likely
to account for the deficiencies of the resulting model. The addition
of quark loops necessary to remove the quenching introduces meson loops
when the theory is viewed at the hadron level \cite{Durand},
and the effect of these loops
on the baryon moments must be considered. We develop our approach to
loop effects in Sec.\ III, where we define the chiral couplings we will
use for low momentum meson-baryon and electromagnetic interactions
in a calculation based on heavy-baryon perturbation theory \cite{HBPT,HBChPT}.
We also introduce the form factor needed to account for the extended
structure of the baryons and mesons viewed as composite particles. We present
our final expression for the octet moments, the results of our numerical
analysis, and our conclusions in Sec.\ IV. Finally, we give the
detailed results needed to evaluate the formal expressions for the moments
in two appendices.

\section{Baryon moments in a QCD-based quark model}
\subsection{Background}
\label{subsec:background}

In previous work \cite{Durand}, we analyzed the QM for the baryon moments
in the context of QCD. Our approach was based on the
work of Brambilla {\em et al.} \cite{brambilla}, who derived the interaction
potential and wave equation for the valence quarks in a baryon
from QCD using a Wilson-line construction. Their basic idea was to construct a
Green's function for the propagation  of a gauge-invariant combination of
quarks joined by path ordered Wilson-line factors
\begin{equation}
U=P\exp ig\!\int\! A_g\cdot dx, \label{wilsonline}
\end{equation}
where $A_g$ is the color gauge field. The Wilson lines sweep out a
three-sheeted world sheet of the form shown in Fig.1 as the quarks move from
their initial to their final configurations.

By making an expansion in powers of $1/m_q$  using the Foldy-Wouthuysen
approximation, and considering only forward
propagation of the quarks in time,
Brambilla {\em et al.} were able to derive a Hamiltonian and Schr\"{o}dinger
equation for the quarks, with an interaction which involves an average over the
gauge field. That average was performed using the minimal surface approximation
in which fluctuations in the world sheet are ignored,
and the geometry is chosen to minimize the total area of the world sheet
subject to the motion of the quarks. The short-distance QCD
interactions were taken into account explicitly. Finally, the kinetic terms
could be resummed. The result of this construction is an effective Hamiltonian
\cite{brambilla} to be used in a semirelativistic Schr\"{o}dinger
equation $H\Psi=E\Psi$,
\begin{eqnarray}
H &=& \sum_i\sqrt{p_i^2+m_i^2}+\sigma(r_1+r_2+r_3)-\sum_{i<j}\frac{2}{3}
\frac{\alpha_{\rm s}}{r_{ij}} \nonumber \\
&&-\frac{1}{2m_1^2}\frac{\sigma}{r_1}{\bf s}_1
\cdot ({\bf r}_1\times{\bf p}_1)
+\frac{1}{3m_1^2}{\bf s}_1\cdot[({\bf r}_{12}\times{\bf p}_1)
\frac{\alpha_{\rm s}}{r_{12}^3} + ({\bf r}_{13}\times{\bf p}_1)
\frac{\alpha_{\rm s}}{r_{13}^3}] \nonumber\\
&&-\frac{2}{3}\frac{1}{m_1m_2}\frac{\alpha_{\rm s}}{r_{12}^3}
{\bf s}_1\cdot{\bf r_{12}}\times{\bf p}_2
-\frac{2}{3}\frac{1}{m_1m_3}\frac{\alpha_{\rm s}}{r_{13}^3}
{\bf s}_1\cdot{\bf r_{13}}\times{\bf p}_3+\cdots.
\label{hamiltonian}
\end{eqnarray}
Here ${\bf r}_{ij}={\bf x}_i-{\bf x}_j$ is the
separation of quarks $i$ and $j$, $r_i$ is the distance of quark $i$ from
point at which the sum $r_1+r_2+r_3$ is minimized, and ${\bf p}_i$ and
${\bf s}_i$ are the momentum and spin operators for quark $i$.
The parameter $\sigma$
is a ``string tension'' which specifies the strength of the long range
confining interaction, and $\alpha_{\rm s}$ is the strong coupling.
The terms hidden in the ellipsis include tensor and spin-spin
interactions which will not play a role in the
analysis of corrections to the moment operator, and the terms that
result from permutations of the particle
labels. The full Hamiltonian is given in \cite{brambilla}.
This Hamiltonian, including the terms omitted here,
gives a good description of the
baryon spectrum as shown by Carlson, Kogut, and Pandharipande \cite{kogut}
and by Capstick and Isgur \cite{isgur}, who proposed it on the basis of
reasonable physical arguments, but did not give formal derivations from
QCD.

The presence of the quark momenta ${\bf p}_i$ in the Thomas-type spin-dependent
interactions in Eq.\ (\ref{hamiltonian}) suggests that new contributions to
the magnetic moment operator could arise in a complete theory through
the minimal substitution
\begin{equation}
{\bf p}_i\rightarrow {\bf p}_i-e_i{\bf A}_{\rm em}(x_i), \label{minimalsub}
\end{equation}
with ${\bf A}_{\rm em}(x_i)$ the electromagnetic
vector potential associated with an
external magnetic field. We have therefore redone the calculation of
Brambilla {\em et al.} with the gauge interaction extended to include
$A_{\rm em}$. After reorganizing the calculation to keep the presence
of $A_{\rm em}$ explicit
throughout, and then expanding to first order in $e_qA_{\rm em}$ in the baryon
rest frame with
\begin{equation}
{\bf A}_{\rm em}=\frac{1}{2}{\bf B}\times{\bf x}_q,
\quad {\bf B}={\rm constant,} \label{magneticpotential}
\end{equation}
we can identify the modified magnetic moment operator through the relation
\begin{equation}
\Delta H=-\mbox{\boldmath$\mu$}\cdot{\bf B}. \label{deltaH}
\end{equation}

The new moment operator, $\mbox{\boldmath$\mu$}=\mbox{\boldmath$\mu$}^{\rm
(QM)}+\Delta\mbox{\boldmath$\mu$}^{QM}$, involves
the leading corrections to the
quark-model operator associated with the binding interactions.
$\Delta\mbox{\boldmath$\mu$}$ {\em can}, in fact, be read off from the
terms in Eq.\ (\ref{hamiltonian}) which depend on both the quark spins
and momenta by making the minimal substitution in Eq.\ (\ref{minimalsub}).
For example, the term which depends on ${\bf s}_1\cdot{\bf r}_{12}\times
{\bf p}_1$ gives an extra contribution
\begin{equation}
\frac{e_1}{6m_1^2}{\bf x}_1\times({\bf s}_1\times{\bf r}_{12})\frac{\alpha_
{\rm s}}{r_{12}^3} \label{deltamu1}
\end{equation}
to $\mbox{\boldmath$\mu$}_1$. There are also possible orbital contributions
to the moments because the Hamiltonian mixes states with nonzero orbital
angular momenta with the ground state. These contributions proved
to be negligible \cite{Durand}.

The moment of a baryon $b$ is now given by
\begin{equation}
\mu_b=\sum_{q}(\mu_q+\Delta\mu_q^b)\langle\sigma_{q,z}\rangle_b
=\mu_b^{QM}+\sum_q\Delta\mu_q^b\langle\sigma_{q,z}\rangle_b,
\label{QMmufinal}
\end{equation}
where the sum is over the quarks in the baryon and
we have quantized along $\bf B$, taken along the $z$ axis. The spin
expectation values are to be calculated in the baryon ground state.

Note that the
correction $\Delta\mu_q^b$ to the  moment of quark $q$
depends on the baryon $b$ in which it appears. The final baryon moments depart
from the quark model pattern only when
the ratios $\Delta\mu_q^b/\mu_q$ differ in different baryons.

\subsection{Results for the octet baryons}
\label{subsec:QMresults}

The general result for the moment operator given above
can be simplified considerably for the $L=0$
ground state baryons. The absence of any orbital angular momentum
allows us to integrate immediately over angles. Furthermore,
two quarks always have the same mass in each octet or decuplet baryon,
so appear symmetrically in the spatial part of a flavor-independent
wave function. We will label these quarks 1 and 2, let $m_1=m_2=m$,
and take 3 as the odd-mass quark if there is one.
With this labeling, the spatial matrix elements of the new operators
can be reduced to a small set, and the correction terms become
\begin{eqnarray}
\Delta\mu_1^{QM,b} &=&\mu_1\left[
\frac{\epsilon+\epsilon'}{2m}+\frac{1}{e_1}\left(
\frac{e_2}{m}\epsilon+\frac{e_3}{m_3}\tilde{\epsilon}\right)
-\frac{\Sigma+\Sigma'}{2m}\right],\quad  \nonumber\\
\Delta\mu_2^{QM,b}&=& \mu_2\left[
\frac{\epsilon+\epsilon'}{2m}+\frac{1}{e_2}\left(
\frac{e_1}{m}\epsilon+\frac{e_3}{m_3}\tilde{\epsilon}\right)
-\frac{\Sigma+\Sigma'}{2m}\right],\,\nonumber\\
\Delta\mu_3^{QM,b} &=&
\mu_3\left[\frac{\tilde{\epsilon}}{m_3}+\frac{1}{e_3}\frac{e_1+e_2}{m}
\epsilon'-\frac{\widetilde{\Sigma}}{m_3}\right],\label{deltas}
\end{eqnarray}
where the $\epsilon$'s and $\Sigma$'s are ground state radial matrix elements,
\begin{eqnarray}
\epsilon &=& \langle\frac{2\alpha_s}{9}\frac{{\bf r}_{12}\cdot{\bf x}_1}
{r_{12}^3}\rangle_{_{\scriptstyle b}} =
\langle\frac{\alpha_s}{9}\frac{1}{r_{12}}\rangle_
{_{\scriptstyle b}}\,,
\nonumber \\
\epsilon' &=& \langle\frac{2\alpha_s}{9}\frac{{\bf r}_{23}\cdot{\bf x}_2}
{r_{23}^3}\rangle_{_{\scriptstyle b}} = \langle\frac{2\alpha_s}{9}\frac{{\bf
r}_{13}
\cdot{\bf x}_1}
{r_{31}^3}\rangle_{_{\scriptstyle b}}\,, \nonumber \\
\tilde{\epsilon} &=& \langle\frac{2\alpha_s}{9}\frac{{\bf r}_{31}\cdot{\bf
x}_3}
{r_{31}^3}\rangle_{_{\scriptstyle b}} = \langle\frac{2\alpha_s}{9}\frac{{\bf
r}_{32}
\cdot{\bf x}_3}
{r_{23}^3}\rangle_{_{\scriptstyle b}}\,,
\label{epsilons}
\end{eqnarray}
and
\begin{eqnarray}
\Sigma &=& \frac{\sigma}{6}\langle\frac{{\bf r}_{12}\cdot{\bf x}_1}{r_{12}}
\rangle_{_{\scriptstyle b}} = \frac{\sigma}{12}\langle
r_{12}\rangle_{_{\scriptstyle b}}\,,\nonumber\\
\Sigma' &=& \frac{\sigma}{6}\langle\frac{{\bf r}_{23}\cdot{\bf x}_2}{r_{23}}
\rangle_{_{\scriptstyle b}} = \frac{\sigma}{6}\langle\frac{{\bf
r}_{31}\cdot{\bf x}_1}{r_{31}}
\rangle_{_{\scriptstyle b}} \,,\nonumber\\
\widetilde{\Sigma} &=& \frac{\sigma}{6}\langle\frac{{\bf r}_{31}
\cdot{\bf x}_3}{r_{31}}\rangle_{_{\scriptstyle b}}
= \frac{\sigma}{6}\langle\frac{{\bf r}_{32}\cdot{\bf x}_3}{r_{23}}
\rangle_{_{\scriptstyle b}}\,.
\label{sigmas}
\end{eqnarray}
In writing these results, we have made the approximation $r_1+r_2+r_3\approx
\frac{1}{2}(r_{12}+r_{23}+r_{31})$, known to be reasonably accurate for the
ground state baryons \cite{kogut}, and used the corresponding Thomas spin
interaction.

We have evaluated the the radial matrix elements above using Gaussian
wave functions obtained in  a variational calculation of the ground
state energies for the Hamiltonian in Eq.\ (\ref{hamiltonian}). The results
are given in Table \ref{table:matrix_elements} for
$\alpha_s=0.39$ and $\sigma=0.18\ {\rm GeV}^2$, values taken from
fits to the baryon spectrum \cite{kogut,isgur} using the same Hamiltonian.
A refitting of the octet moments using $\mu_u$ and
$\mu_s$ as adjustable parameters in the QM contribution to the complete
expression in Eq.\ (\ref{QMmufinal}) gives only a slight improvement in
the results, with an incorrect pattern in the corrections relative to those
needed.

\section{LOOP CORRECTIONS TO THE MOMENTS}

Given the failure of the binding corrections to eliminate the deficiencies
in the QCD-based quark model, we have reexamined the approximations used
in its derivation \cite{Durand}.
A key element in the Wilson-loop construction was the
use of the quenched approximation in which all internal quark loops are
omitted. As discussed elsewhere\cite{Durand}, it appears that this
approximation is the one most likely to account for the difficulty in
reproducing the measured moments. In particular, the
introduction of quark loops allows
meson loops to appear, and these are known to affect magnetic moments.
The first step in the improvement of the model is therefore the introduction
of meson loop corrections to the QM moments. The relevant one-loop
Feynman diagrams at the hadronic level are shown in Figs.\ \ref{fig:mesonloop}
and \ref{fig:baryonloop}. Since the
ground state $L=0$ octet and decuplet baryons only differ in their internal
spin configurations in the simple QM and the octet-decuplet mass
differences are small, we must include both sets of baryon states in the
calculation to get a consistent theory. However, we include only the
pseudoscalar mesons.

The diagrams in Fig.\ \ref{fig:mesonloop} are independent of the input magnetic
moments of the baryons, but modify the final moments.
In contrast, the diagrams in Fig.\ \ref{fig:baryonloop} involve the
octet and decuplet moments and the octet-decuplet transition moment
directly. We will specify these in terms of the QM.

We will calculate the contributions of the diagrams in Figs.\
\ref{fig:mesonloop} and \ref{fig:baryonloop}
using heavy baryon perturbation theory (HBPT) in which the baryon masses
are assumed to be large compared to the typical scale set by the internal
momenta. It will be essential in this respect to take the extended, composite
structure of the baryons into account, since this extended structure naturally
limits the the momenta that can be absorbed by the baryon as a whole.
The resulting momenta are small on the average, well below usual chiral
cutoff at $\approx 1$ GeV, and it is therefore reasonable
to use the lowest order
chiral couplings to describe the resulting the low momentum, long distance
interactions of the mesons and baryons.\footnote{The difficulties with
ChPT in the present context result from the treatment of the baryons and
mesons as point particles in HBChPT. This leads to divergences in loop
integrals. Different regularization schemes lead to different results for
the integrals, for example, in dimensional \cite{krause} and
momentum \cite{donoghue} regularization,
with the ambiguities being lumped into the new couplings that
must be introduced along with the loop diagrams. The present theory has a
natural cutoff at fairly low momentum imposed by the extended structure of the
hadrons in the QM, and no new couplings appear.}
We will discuss these ingredients
separately in the following subsections.

\subsection{Heavy baryon chiral couplings}

Heavy baryon perturbation theory was developed in Ref.\ \cite{HBPT} and
extended to the chiral context in Ref.\
\cite{HBChPT}. It has been used to study a number of hadronic processes
at momentum transfers much less than 1 GeV. The key ideas in HBPT
involve the replacement of the
momentum $p^\mu$ of a nearly on-shell baryon by its on-shell momentum
$m_Bv^\mu$ plus a small additional momentum $k^\mu$, $p=m_Bv+k$, and the
replacement of the baryon field operator $B(x)$ by an operator $B_v(x)$
constructed to remove the free momentum dependence in the Dirac equation,
\begin{equation}
  B_v (x) = e^{i m_B {\not v} v^{\mu} x_{\mu}}B(x) \ .
\end{equation}
In these expressions $v^\mu$ is the on-shell four velocity of the baryon,
and it is assumed that $k\cdot v\ll m_B$. The perturbation expansion then
involves modified Feynman rules and an expansion in powers of $k/m_B$. We
will work to leading order in this expansion.

The chiral Lagrangian for the modified baryon fields depends on the usual
matrix of baryon fields, with $B$ replaced by $B_v$, and on the pseudoscalar
pion octet normalized as
\begin{equation}\label{eq:1}
\bbox{\phi} = {1 \over \sqrt{2}} \left(
\begin{array}{ccc}
{\pi^0\over\sqrt{2}} + {\eta\over \sqrt{6}} & \pi^+ & K^+ \\ \pi^- & -{\pi^0
\over \sqrt{2}} + {\eta \over \sqrt{6}} & K^0 \\ K^- & \overline K^0 & - {2\eta
\over \sqrt{6}}
\end{array} \right) \,.
\end{equation}
This couples to the baryon fields at low momenta through the
vector and axial vector currents defined by
\begin{eqnarray}
V_\mu = f^{-2} (\phi\partial_\mu\phi - \partial_\mu\phi \phi)+\cdots  \,, \ \
A_\mu&=& f^{-1}\partial_\mu \phi+\cdots \,, \ \
\end{eqnarray}
where $f \approx 93$ MeV is the meson decay constant. We will retain, as
indicated above, only leading term in the derivative expansion.
The lowest order chiral Lagrangian for octet and decuplet baryons is then
\begin{eqnarray}\label{lag}
{\cal L}_v &=& i \ {\rm Tr}\ \bar B_v\ \left(v\cdot {\cal D}
\right)B_v
+ 2\ D\  {\rm Tr}\ \bar B_v\ S_v^\mu\ \{ A_\mu, B_v \}
+ 2\ F\ {\rm Tr}\  \bar B_v\ S_v^\mu\ [A_\mu, B_v]
\nonumber \\
&&-\ i\ \bar
T_v^{\mu}\ (v \cdot {\cal D}) \  T_{v \mu}
+ \delta \ \bar T_v^{\mu}\ T_{v \mu}
+ {\cal C}\ \left(\bar T_v^{\mu}\ A_{\mu}\ B_v + \bar B_v\ A_{\mu}\
T_v^{\mu}\right){\phantom {f^2 \over 4}}
\nonumber\\
&& +\ 2\ {\cal H}\  \bar T_v^{\mu}\ S_{v \nu}\ A^{\nu}\  T_{v \mu}
+ {\rm Tr}\ \partial_\mu \phi \partial^\mu
\phi +\cdots \, \ \
\end{eqnarray}
where $\delta$ is the decuplet-octet mass difference, and
${\cal D_\mu}= \partial_\mu+[V_\mu,  \; ]$ is the covariant chiral derivative.
$B_v$ is now the matrix of octet baryon fields, and the
Rarita-Schwinger fields $T_v^\mu$ \cite{HBChPT} represent the decuplet
baryons. $D$, $F$, ${\cal C}$, and ${\cal H}$ are the strong interaction
coupling constants. The spin operator $S_v^{\mu}$ is defined in
Ref.\cite{HBChPT}.

The electromagnetic interactions of the mesons, and the convection
current interactions of the baryons, are introduced into the
Lagrangian by making the substitutions
\begin{eqnarray}
{\cal D}_\mu & \mbox{$\rightarrow$} & {\cal D}_\mu+ie{\cal A}_\mu[Q,\ ],
\nonumber\\
\partial_\mu \phi & \mbox{$\rightarrow$} & {\cal D_\mu} \phi=
\partial_\mu\phi+ie{\cal A}_\mu[Q,\phi],
\end{eqnarray}
where ${\cal A}_\mu$ is the photon field.
The baryon moment couplings needed in the graphs in Fig.\ \ref{fig:baryonloop}
do not appear above, and require a separate treatment.

\subsection{Quark model moment couplings}
\label{subsec:QMmoments}

As described above, our starting point for calculating the baryon moments
is the QCD-based quark model. These moments play the role of tree-level
interactions in the loop expansion,
and also appear in the electromagnetic interactions of the internal
baryons in Fig.\ \ref{fig:baryonloop}.
As we noted in \cite{DandH}, the QM baryon moments
can be written in terms of the general SU(3) symmetry breaking operator
in HBChPT given by \cite{Jenetal,Luty,Daietal,Mei,Bos},
\begin{equation}
\label{qmop}
  {\cal L}_{SB} = {\cal L}_0 + {\cal L}_1
\end{equation}
where %
\begin{equation}
\label{chiop}
   {\cal L}_0 = {e \over 4 m_N}\Big(
        \mu_D{\rm Tr}\,\bar{B_v} F_{\mu\nu}\sigma^{\mu\nu} \{Q,B_v \} \,
        + \mu_F{\rm Tr}\,\bar{B_v} F_{\mu\nu}\sigma^{\mu\nu} [Q,B_v]\Big)\,\;,
\end{equation}
is the leading order moment operator in chiral perturbation
theory parametrized by $\mu_D$ and $\mu_F$ \cite{Coleman}, and
\begin{eqnarray}
\label{sbop}
   {\cal L}_1 &=& {e \over 4 m_N}F_{\mu\nu}\Big(
        c_1{\rm Tr}\,\bar{B_v}{\cal M} Q\sigma^{\mu\nu}B_v +
        c_2{\rm Tr}\,\bar{B_v}Q\sigma^{\mu\nu}B_v{\cal M} +
        c_3{\rm Tr}\,\bar{B_v}\sigma^{\mu\nu}B_v{\cal M}Q  \nonumber \\
    &+& c_4{\rm Tr}\,\bar{B_v}{\cal M}\sigma^{\mu\nu}B_vQ +
        c_5{\rm Tr}\,\bar{B_v}\sigma^{\mu\nu}B_v {\rm Tr}\,{\cal M}Q \, \Big)
\end{eqnarray}
contains the new couplings $c_1,\ldots,c_5$ introduced along with the
counterterms that are necessary at one loop
\cite{Jenetal,Luty,Daietal,Mei,Bos}. Here
$Q={\rm diag}(2/3,-1/3,-1/3)$ is
the quark charge matrix, and ${\cal M}={\rm diag}(0,0,1)$ is proportional
to the mass matrix used to introduce SU(3) breaking through the strange-quark
mass $m_s$. The addition to ${\cal L}_{SB}$ of one
further coupling which is second order in $m_s$,\footnote{Since ${\cal MQ}=
{\cal QM}=-\frac{1}{3}{\cal M}$, the rearrangements of the factors
${\cal MQ}$ and $\cal M$ in Eq.\ (\ref{eighth_coupling}) give no new
contributions to ${\cal L}_2$, and the form given is the only new invariant.}
\begin{equation}
{\cal L}_2= {e d\over 4 m_N}Tr\bar{B}_v{\cal M}Q
\sigma^{\mu\nu}F_{\mu\nu}B_v{\cal M}
\label{eighth_coupling}
\end{equation}
gives a complete
basis for the description of the octet moments \cite{DandH}.

The particular choice $\mu_F=2\mu_D/3=\mu_u$ and $c_1=-5\Delta$, $c_2=0$,
$c_3=-\Delta$, $c_4=0$, and $c_5=\Delta$, $d=0$
for the parameters above, with $\Delta=(2\mu_s+\mu_u)/2$,
gives us the QM moments

\begin{eqnarray} \label{qml}
 \mu_p^{(QM)} = {3 \over 2} \mu_u \ , \ \ \ \
 \mu_{\Sigma^+}^{(QM)} = {4 \over 3} \mu_u - {1 \over 3} \mu_s \ , \ \ \
 \mu_{\Xi^0}^{(QM)} = {4 \over 3} \mu_s - {1 \over 3} \mu_u \ , \nonumber \\
 \mu_n^{(QM)} = - \mu_u \ , \ \ \ \
 \mu_{\Sigma^0}^{(QM)} = {1 \over 3} (\mu_u - \mu_s)  \ , \ \ \
 \mu_{\Xi^-}^{(QM)} = {1 \over 6} \mu_u + {4 \over 3} \mu_s  \ , \\
 \mu_{\Lambda}^{(QM)} = \mu_s \ , \ \ \ \ \
 \mu_{\Sigma^-}^{(QM)} = -{2 \over 3} \mu_u - {1 \over 3} \mu_s \ , \ \
 \mu_{\Lambda\Sigma^0}^{(QM)}
    = {\sqrt{3} \over 2}\mu_u  \ . \ \ \ \ \ \ \nonumber
\end{eqnarray}

Since the one loop corrections involve both intermediate octet and decuplet
baryon states, we also need the Lagrangians for the decuplet magnetic moment
couplings and the octet-decuplet magnetic moment transitions for the QM.
These are given respectively by
\begin{equation} \label{decupmm}
{\cal L}^{(3/2)} = -i{3 e \over 2 m_N}
       \bar T_{v ikl}^{\mu} {\widehat Q}^i_j T_{v }^{\nu jkl} F^{\mu \nu} \ ,
\end{equation}
and
\begin{equation}
{\cal L}^{(od)} = -i{2 e \over m_N} F^{\mu \nu}(\epsilon_{ijk}
     {\widehat Q}_l^i \bar B_{vm}^j S_v^\mu T_v^{\nu klm} + h.c ) \ ,
\end{equation}
where $i$, $j$, $k$, $l$, and $m$ are SU(3) flavor indices and $\widehat
Q=\rm{diag} (\mu_u, -\mu_u/2, \mu_s)$.

\subsection{ Meson wave function effects: The form factor}

The baryons in the QM are composite states, and can absorb only limited recoil
momentum while remaining in their ground states. This must be taken into
account in a dynamical model. In the absence of a detailed theory of the
interactions of composite mesons and baryons, we will model the wave
function effects using a form factor at each meson-baryon vertex.
In keeping with the heavy baryon picture, we will define the form factor in
the rest frame of the baryon where it can depend only on $\bf k^2$, the square
of the three momentum of the meson. A form factor of this type automatically
repects crossing since a change $\bf k\rightarrow-k$ corresponding to
a shift of a meson between the initial and final states does not
change the form factor.

We have chosen for simplicity to use a one-parameter form factor
\begin{equation} \label{formf}
F({\bf k}) = { \lambda^2 \over {\lambda^2 + {\bf k}^2}}
\end{equation}
normalized at chiral limit, ${\bf k}=0$. The parameter
$\lambda$ characterizes the natural momentum
scale, expected to be much below 1 GeV.
With the introduction of this form factor, all the diagrams in
Figs.\ \ref{fig:mesonloop} and \ref{fig:baryonloop} are finite and no arbitrary
subtractions
need to be introduced into the theory.

The evaluation of momentum integrals in the presence of the form factor
is straightforward, but involves some new elements. First, the form factor
(\ref{formf}) is rewritten in covariant form as
\begin{equation}\label{formfr}
F(k,v)={- \lambda^2  \over k^2 -(k \cdot v)^2 - \lambda^2} \ .
\end{equation}
The factors in the denominator of a one-loop integrand
with integration variable
$k$ can be combined using the Feynman parametrization formula to obtain an
expression of the form
\begin{equation}\label{deno}
k^2 + \alpha (k \cdot v)^2 + \beta (k \cdot v) + \gamma \ .
\end{equation}
Here $\alpha$, $\beta$, and $\gamma$ are quantities independent of
the loop momentum $k$, and the photon momentum $q$ has been set equal
to zero in the denominators. A change of the variable of integration to
\begin{equation}\label{transf}
k^{'} = k + [\pm\sqrt{1+\alpha}-1]\, v (k \cdot v)
\end{equation}
then brings Eq.(\ref{deno}) to the standard form
\begin{equation}\label{deno1}
k^{'2} \pm {\beta \over \sqrt{1+\alpha}}(k^{'} \cdot v) + \gamma \ ,
\end{equation}
and the loop integral is easily evaluated. Note that the Jacobian of the
transformation of variables in Eq.(\ref{transf}) is $1/\sqrt{1+\alpha}$.

\section{Baryon magnetic moments}
\subsection{Expressions for the octet baryon moments}

We can now give our expressions for the baryon magnetic moments
including the loop corrections from the diagrams shown in Figs.\
\ref{fig:mesonloop}
and \ref{fig:baryonloop}. In units of nuclear magnetons, the moment of baryon
$i$ is

\begin{equation}
\lambda_i = \mu_i^{(0)} + \mu_i^{(1/2)} + \mu_i^{(3/2)}\ ,
\end{equation}
where the leading term $\mu_i^{(0)}$ includes the QM moments
plus the corrections $\Delta\mu_i^{QM}$ obtained in the QCD-based QM, while
$\mu_i^{(1/2)}$ and $\mu_i^{(3/2)}$ are the contributions from the one-loop
graphs with intermediate octet and decuplet
baryon states, respectively. We find that

\begin{eqnarray}
\mu_i^{(0)}&=&\alpha_i + \Delta\mu_i^{QM} \ , \\
\mu_i^{(1/2)} &=& \sum_{X=\pi,K}- { m_N \over {24\pi f^2}}
        {\lambda^4 \over (\lambda+m_X)^3} \beta_i^{(X)}  \,
\qquad  \nonumber \\
       &+& \sum_{X=\pi,K,\eta}{1 \over {16\pi^2 f^2}}
         (\gamma_i^{1(X)}-2\lambda_i^{(X)}\alpha_i )
 L_0( m_X, \lambda) \ , \label{mu12}
\end{eqnarray}
and
\begin{eqnarray} \label{mu32}
\mu_i^{(3/2)} &= & \sum_{X=\pi,K} {m_N \over 8\pi f^2} \,
      \widetilde F(m_X ,   \delta, \lambda)\tilde \beta_i^{(X)}  \nonumber \\
     &+& \sum_{X=\pi,K,\eta}{1 \over 32\pi^2 f^2}
  \Big [ \ ( \tilde \gamma_i^{1(X)}
  -2 \tilde \lambda_i^{(X)}\alpha_i ) L_1(m_X , \delta, \lambda) \ +
    \tilde \gamma_i^{2(X)} L_2(m_X , \delta, \lambda) \Big ]  \ ,
\end{eqnarray}
where $\alpha_i=\mu_i^{(QM)} $. The coupling coefficients $\beta_i^{(X)}$,
$\tilde \beta_i^{(X)}$, $\lambda_i^{(X)}+\tilde \lambda_i^{(X)}$ are identical
to those in Ref. \cite{Jenetal}, and will not be given here.\footnote{
The tadpole graph 2b in Ref. \cite{Jenetal} is absent in our model.
Since the chiral symmetry is broken in the QM, we do not include chiral
corrections to the baryon moment operators as in that reference.
The elementary ${\bar B}B\phi\phi{\cal A}_\mu$ vertex
connected with spin operator therefore does not exist.}
The remaining coefficients $\gamma_i^{1(X)}$, $\tilde \gamma_i^{1(X)}$,
and $\tilde \gamma_i^{2(X)}$ depend explicitly on $\mu_u$ and $\mu_s$, and have
not appeared elsewhere. We list those coefficients in Appendix A.

To connect the various terms to the loop graphs in Figs.\ \ref{fig:mesonloop}
and \ref{fig:baryonloop}, we note that $\beta_i^{(X)}$, $\tilde \beta_i^{(X)}$,
$\gamma_i^{1(X)}$, $\tilde \gamma_i^{1(X)}$, $\tilde \gamma_i^{2(X)}$,
$\lambda_i^{(X)}$, and $\tilde \lambda_i^{(X)}$  are the coefficients of the
graphs 2a, 2b, 3a, 3b, 3c (or 3d), 3e and 3f, respectively.

The expressions for the functions $ L_0(m_X,\lambda)$, $\widetilde
F(m_X,\delta,\lambda)$,
$ L_1(m_X ,\delta,\lambda)$, and $L_2(m_X ,\delta,\lambda)$
are given in the Appendix B. These functions result from the loop integrations,
and depend on the meson masses, the decuplet-octet mass difference
$\delta$, and the natural wave function cutoff $\lambda$. We emphasize that
the loop integrations are all finite, and that the wave function parameter
$\lambda$ sets the natural momentum scale in graphs that are divergent for
point particles.

\subsection{Fits to the data}

We have used the foregoing expressions to fit the experimental baryon magnetic
moments. It is worth emphasizing that all the parameters that appear
except for $\mu_u$, $\mu_s$, and $\lambda$ are known independently.
We evaluated the corrections $\Delta\mu_i^{QM}$ given in
Eqs.\ (\ref{QMmufinal}) and (\ref{deltas}) from the QCD-based
QM using the matrix elements given in Table \ref{table:matrix_elements} and
and quark masses $m_u=m_d=0.343$ GeV and $m_s=0.539$ GeV chosen
to give the best fit to the octet magnetic moments in the naive QM. The strong
interaction couplings $F$, $D$, and $\cal C$ were chosen to
satisfy the SU(6) relations $ F=2D/3,\ {\cal C}=-2D$ expected in the QM, with
the values $F=0.5, D=0.75$, and ${\cal C}= -1.5$ chosen so that $F+D \approx
|g_A/g_V|=1.26$. The decuplet coupling $\cal H$ does not appear.
The decuplet-octet mass difference was taken as
$\delta=250$ MeV; the results are fairly insensitive to this choice.
Finally, we used the values $f_\pi=93$ MeV and $f_K=f_\eta=1.2 f_\pi$
\cite{Jenetal}.

The results of an equal weight least square fit to the seven well-measured
octet moments using the three free parameters $\mu_u$, $\mu_s$, and $\lambda$
are summarized in \ref{table:moments}.
We find root-mean-square deviation of the predicted values from the observation
of 0.055 $\mu_N$. This is a substantial improvement on the naive quark model
which gives an average deviation of 0.12 $\mu_N$. The transition moments
$\mu_{\Sigma^0\Lambda}$ was not included in the fit,
but was left as a prediction. The result obtained, $\mu_{\Sigma^0\Lambda}=
1.559\ \mu_N$, differs from the experimental value $\mu_{\Sigma^0\Lambda}=
1.610\pm 0.08\ \mu_N$, by $0.05\ \mu_N$, a value within the experimental
uncertainty and one that does not affect the overall fit.

A detailed breakdown of the contributions of the
loop integrals to the fitted magnetic moments of the octet baryons is given
in the \ref{table:loop_details}.
The results in this table shows that the loop contributions are
small in comparison to the leading QM contributions, suggesting reasonably
rapid convergence of the loop expansion. This is in marked contrast to the
results obtained in HBChPT, where the loop contributions calculated
using dimensional regularization are comparable in size to the
leading terms in the chiral expansion \cite{DandH}. It is also clear from the
table that the binding corrections $\Delta\mu_i^{QM}$ found in the QCD-based
derivation of the QM \cite{Durand} are important. It is not possible to
obtain as good a fit to the data as given in the Table \ref{table:moments}
when these are omitted.

A further interesting point involves the importance of the
decuplet intermediate states for the octet moments. It is easy to check
that the contributions from the graphs involving decuplet
states, that is, the sum of the contributions from the graphs 2b, 3b, 3c, 3d,
and 3f, are substantial. For most of the baryons, those contributions are
larger than those from the graphs which involve only the intermediate octet
states. This result is insensitive to the value used for the octet-decuplet
mass difference. We conclude that the decuplet must be regarded as
a set of light baryon states as in the QM, and not, as in \cite{Mei}
as a set of heavy states.
The present method for obtaining the loop corrections to the QM moments
should therefore apply to the decuplet moments as well. That calculation
has been carried out elsewhere \cite{Ha} using the same values of $\mu_u$,
$\mu_s$, and $\lambda$ as obtained here, and gives a prediction
$\mu_{\Omega^-}=-1.97\ \mu_N$ in striking agreement with the measured
value $-2.02\pm 0.05\ \mu_N$, and much better than the prediction of the
naive QM, $\mu_{\Omega^-}=-1.74\ \mu_N$.

\subsection{Conclusions and prospects}

In this paper, we have considered the one-loop corrections to the octet baryon
magnetic moments in the QCD-based QM. It is necessary to include these as
a first step in getting away from the quenching of internal quark loops
used in the derivation of the QM.

Our approach to the calculation is based on HBPT. We
use the derivative couplings for the low-momentum interactions of the
pseudoscalar mesons with the baryons favored by ChPT,
and a form factor to characterize the composite structure of baryons. The loop
diagrams are all finite, and no counterterms need to be introduced
to absorb divergences. We are, in fact, making a dynamical calculation
of the extra couplings or counterterms
encountered in HBChPT in the sense that our expressions
for the eight octet moments can be parametrized exactly \cite{DandH} in terms
of the eight chiral couplings defined in Eqs.\ (\ref{chiop}), (\ref{sbop}) and
(\ref{eighth_coupling}).

The results from our fits to the octet baryon moments using the QCD QM
with loop corrections are very good, with an average deviation of the
theoretical moments from theory of $0.05\ \mu_N$, significantly better
than the QM results. The contribution from each loop graph is small
compared to the leading terms, suggesting convergence of the loop
expansion. The parameter $\lambda$ which sets the momentum scale in the
meson-baryon interaction or wave function is about $400$ MeV, a value
consistent with the expectations deduced from the observed transverse momentum
distributions in pion production. This value
is closer to the kaon mass than to the pion mass. As a result, the wave
function effects suppress the short-distance contributions from kaon and
$\eta$-loops, but affect the more reliable long-distance part of the pion
loop contributions relatively little. We conclude that it is crucial
to take the effects of compositeness into account if one is to have
a controllable perturbation theory for hadrons at low momentum.

A question which arises at this point concerns the extent to which the
theory can be further improved by the inclusion of higher-mass mesons and
baryons. The contributions of the ground state vector mesons would be
expected to be as important as the contributions from the pseudoscalar
mesons except for the suppression of high mass intermediate states by
the momentum cutoff imposed by the wave functions. The contributions of
higher mass baryons in the intermediate states are also suppressed.
However, there are low-mass multiparticle intermediate states such as
those with one baryon and two pions that could be important.
A possible approach to the estimation of their effects is through the use
of the sideways or mass dispersion relations proved by Bincer \cite{Bincer}.

We remark, finally, that we believe that this work
demonstrates the importance
of getting beyond the quenched approximation in lattice QCD if one
is to understand the finer details of hadron structure from first
principles.

\acknowledgments
This work was supported in part by the U.S. Department of Energy under
Grant No.\ DE-FG02-95ER40896.
One of the authors (LD) would like to thank the Aspen Center for Physics
for its hospitality while parts of this work were done.

\appendix
\section{The coupling coefficients}
\setcounter{equation}{0}
In this appendix, the coupling coefficients are explained and presented
explicitly. For simplicity, the superscript $(X)$ is suppressed. Our $\beta_i$,
and $\tilde \beta_i$ are identical, respectively, to the coefficients
$\beta_i$  and $\beta_i'$ in Ref. \cite{Jenetal}. The sum of our coefficients
$\lambda_i$ and $\tilde \lambda_i$ is equal to the coefficient
$\bar{\lambda}_i$
defined in \cite{Jenetal},
\begin{equation}
 \label{apa1}
\lambda_i+\tilde \lambda_i = \bar{\lambda}_i.
\end{equation}
Since the couplings associated with graphs which involve only
octet intermediate states are independent of those for graphs which
involve decuplet intermediate states, it is easy to separate $\lambda_i$
and $\tilde \lambda_i$ from the combined coefficient given in \cite{Jenetal}.

The values of the coupling factors $\gamma_i^{1}$ evaluated from the graphs
in Fig.\ 3a are

\begin{eqnarray} \label{gam1p}
 \gamma_p^{1(\pi)} &=& {1 \over 4}(D+F)^2 \mu_u, \nonumber \\
 \gamma_n^{1(\pi)} &=& - (D+F)^2 \mu_u, \nonumber \\
 \gamma_{\Lambda}^{1(\pi)} &=& {2 \over 3} D^2 (\mu_s-\mu_u), \nonumber \\
 \gamma_{\Sigma^+}^{1(\pi)} &=& {2 \over 3}[(3DF - 5F^2) \mu_u + (2F^2-D^2)
\mu_s], \nonumber \\
 \gamma_{\Sigma^0}^{1(\pi)} &=& {2 \over 3} [-2F^2\mu_u + (2F^2-D^2)\mu_s], \\
 \gamma_{\Sigma^-}^{1(\pi)} &=&{2 \over 3}[(-3DF + F^2) \mu_u + (2F^2-D^2)
\mu_s ], \nonumber \\
 \gamma_{\Xi^0}^{1(\pi)} &=& -2(F-D)^2\mu_s, \nonumber \\
 \gamma_{\Xi^-}^{1(\pi)} &=& {1 \over 4} (F-D)^2(\mu_u - 8\mu_s), \nonumber \\
  \gamma_{\Lambda\Sigma^0}^{1(\pi)} &=& {1 \over \sqrt{3}}D(4F-D)\mu_u,
\nonumber
\end{eqnarray}
for the pion loops,
\begin{eqnarray} \label{gam1k}
 \gamma_p^{1(K)} &=& (F-D)(D-3F)\mu_u +
   ({1 \over 3}D^2 - 2DF -F^2)\mu_s , \nonumber \\
 \gamma_n^{1(K)} &=& 2F(F-D)\mu_u +
   ({1 \over 3}D^2 - 2DF -F^2)\mu_s, \nonumber \\
 \gamma_{\Lambda}^{1(K)} &=& -{1 \over 18}(D^2+12DF+9F^2)\mu_u
    - {4 \over 9}(D-3F)^2\mu_s, \nonumber \\
 \gamma_{\Sigma^+}^{1(K)} &=& -{7 \over 6}(D^2-{22 \over 7}DF+F^2)\mu_u -
     {4 \over 3}(D+F)^2\mu_s, \nonumber \\
 \gamma_{\Sigma^0}^{1(K)} &=& -{1 \over 6}(D^2-4DF+F^2)\mu_u -
     {4 \over 3}(D+F)^2\mu_s, \\
 \gamma_{\Sigma^-}^{1(K)} &=&{5 \over 6}(D^2-{14 \over 5}DF+F^2)\mu_u -
     {4 \over 3}(D+F)^2\mu_s, \nonumber \\
 \gamma_{\Xi^0}^{1(K)} &=& -2D(D+F)\mu_u +
      {1 \over 3}(D^2+6DF-3F^2)\mu_s, \nonumber \\
 \gamma_{\Xi^-}^{1(K)} &=& (D^2-F^2)\mu_u +
      {1 \over 3}(D^2+6DF-3F^2)\mu_s  \ , \nonumber \\
 \gamma_{\Lambda\Sigma^0}^{1(K)} &=& {1 \over 2\sqrt{3}}(3D^2+4DF-9F^2)\mu_u
\nonumber
\end{eqnarray}
for the kaon loops, and
\begin{eqnarray} \label{gam1e}
 \gamma_p^{1(\eta)} &=& -{1 \over 4}(3F-D)^2 \mu_u, \nonumber \\
 \gamma_n^{1(\eta)} &=& {1 \over 6} (3F-D)^2 \mu_u, \nonumber \\
 \gamma_{\Lambda}^{1(\eta)} &=& -{2 \over 3} D^2\mu_s, \nonumber \\
 \gamma_{\Sigma^+}^{1(\eta)} &=& -{2 \over 9}D^2(4\mu_u-\mu_s), \nonumber \\
 \gamma_{\Sigma^0}^{1(\eta)} &=&-{2 \over 9}D^2( \mu_u-\mu_s),  \\
 \gamma_{\Sigma^-}^{1(\eta)} &=&{2 \over 9}D^2( 2\mu_u+\mu_s), \nonumber \\
 \gamma_{\Xi^0}^{1(\eta)} &=& {1 \over 18} (D+3F)^2(\mu_u-4\mu_s),
\nonumber \\
\gamma_{\Xi^-}^{1(\eta)} &=& -{1 \over 36} (D+3F)^2(\mu_u+8\mu_s), \nonumber \\
 \gamma_{\Lambda\Sigma^0}^{1(\eta)} &=& {1 \over \sqrt{3}}D^2\mu_u
\nonumber
\end{eqnarray}
for the $\eta$ loops.

The coefficients $\tilde \gamma_i^{1}$ evaluated from
the graphs 3b are given, up to a factor of $-$5${\cal C}^2$/2, by

\begin{eqnarray} \label{tgam1p}
 \tilde \gamma_p^{1(\pi)} &=& -{16 \over 9}\mu_u, \ \ \ \ \ \
 \tilde \gamma_{\Sigma^+}^{1(\pi)} = -{1 \over 27}(5\mu_u+4\mu_s), \ \
 \tilde \gamma_{\Xi^0}^{1(\pi)} = -{4 \over 9} \mu_s ,  \nonumber \\
 \tilde \gamma_n^{1(\pi)} &=&  {4 \over 9} \mu_u , \ \ \ \ \ \ \ \ \ \
 \tilde \gamma_{\Sigma^0}^{1(\pi)} = -{2 \over 27}(\mu_u+2\mu_s), \ \ \ \
 \tilde \gamma_{\Xi^-}^{1(\pi)} = -{1 \over 9}(\mu_u+4\mu_s), \\
 \tilde \gamma_{\Lambda}^{1(\pi)} &=& -{1 \over 3} (\mu_u+2\mu_s), \ \
 \tilde \gamma_{\Sigma^-}^{1(\pi)} = {1 \over 27}(\mu_u-4\mu_s), \ \
 \tilde \gamma_{\Lambda\Sigma^0}^{1(\pi)} = -{2 \over 3\sqrt{3}}\mu_u,
\nonumber
\end{eqnarray}
for the pion loops, by
\begin{eqnarray} \label{tgam1k}
 \tilde \gamma_p^{1(K)} &=& -{1 \over 9}(3\mu_u+2\mu_s), \ \ \
 \tilde \gamma_{\Sigma^+}^{1(K)} = -{2 \over 27}(23\mu_u+4\mu_s), \ \
 \tilde \gamma_{\Xi^0}^{1(K)} = -{1 \over 9}(3\mu_u+14\mu_s) ,  \nonumber \\
 \tilde \gamma_n^{1(K)} &=& {1 \over 9}(\mu_u-2\mu_s), \ \ \ \ \ \
 \tilde \gamma_{\Sigma^0}^{1(K)} = -{1 \over 27}(13\mu_u+8\mu_s), \ \ \
 \tilde \gamma_{\Xi^-}^{1(K)} = {1 \over 9}(\mu_u-14\mu_s), \\
 \tilde \gamma_{\Lambda}^{1(K)} &=& -{1 \over 9}(\mu_u+8\mu_s), \ \ \ \
 \tilde \gamma_{\Sigma^-}^{1(K)} = {4 \over 27}(5\mu_u-2\mu_s), \ \ \ \ \ \ \
 \tilde \gamma_{\Lambda\Sigma^0}^{1(K)} = -{1 \over 3\sqrt{3}}\mu_u, \nonumber
\end{eqnarray}
for the kaon loops, and by
\begin{eqnarray} \label{tgam1e}
 \tilde \gamma_p^{1(\eta)} &=& 0, \ \ \ \ \ \
 \tilde \gamma_{\Sigma^+}^{1(\eta)} = -{2 \over 9}(2\mu_u+\mu_s), \ \ \
 \tilde \gamma_{\Xi^0}^{1(\eta)} = -{2 \over 9}(\mu_u+2\mu_s),  \nonumber \\
 \tilde \gamma_n^{1(\eta)} &=& 0, \ \ \ \ \ \
 \tilde \gamma_{\Sigma^0}^{1(\eta)} = -{1 \over 9}(\mu_u+2\mu_s), \ \ \
 \tilde \gamma_{\Xi^-}^{1(\eta)} = {1 \over 9}(\mu_u-4\mu_s), \\
 \tilde \gamma_{\Lambda}^{1(\eta)} &=& 0 , \ \ \ \ \ \
 \tilde \gamma_{\Sigma^-}^{1(\eta)} = {2 \over 9}(\mu_u-\mu_s), \ \ \ \ \ \ \
 \tilde \gamma_{\Lambda\Sigma^0}^{1(\eta)} = 0,  \nonumber
\end{eqnarray}
for the $\eta$ loops.

The coefficients $\tilde \gamma_i^{2}$ evaluated from
the graphs 3c (or 3d) are given, up to a factor of $2{\cal C}$, by

\begin{eqnarray} \label{tgam2p}
 \tilde \gamma_p^{2(\pi)} &=& -{8 \over 3}(D+F)\mu_u, \ \
 \tilde \gamma_{\Sigma^+}^{2(\pi)} = -{2 \over 9}((3D+5F)\mu_u-8F\mu_s), \ \
 \tilde \gamma_{\Xi^0}^{2(\pi)} = {4 \over 3}(F-D)\mu_s ,  \nonumber \\
 \tilde \gamma_n^{2(\pi)} &=&  {8 \over 3}(D+F) \mu_u , \ \ \ \
 \tilde \gamma_{\Sigma^0}^{2(\pi)} = -{4 \over 9}F(\mu_u-4\mu_s), \ \ \ \
 \tilde \gamma_{\Xi^-}^{2(\pi)} = {2 \over 9}(D-F)(5\mu_u-2\mu_s), \\
 \tilde \gamma_{\Lambda}^{2(\pi)} &=& {2 \over 3}D(\mu_u-4\mu_s) , \ \
 \tilde \gamma_{\Sigma^-}^{2(\pi)} = {2 \over 9}((3D+F)\mu_u+8F\mu_s), \ \
 \tilde \gamma_{\Lambda\Sigma^0}^{2(\pi)} = -{1 \over 3\sqrt{3}}(D+6F)\mu_u
\nonumber
\end{eqnarray}
for the pion loops, by
\begin{eqnarray} \label{tgam2k}
  \tilde \gamma_p^{2(K)} &=& -{4 \over 3}[D\mu_u+(F-D)\mu_s], \nonumber \\
  \tilde \gamma_n^{2(K)} &=& {2 \over 3}[(D+F)\mu_u+2(D-F)\mu_s], \nonumber \\
  \tilde \gamma_{\Lambda}^{2(K)} &=& {2 \over 9}(3F-D)(\mu_u-4\mu_s), \nonumber
\\
 \tilde \gamma_{\Sigma^+}^{2(K)}&=&{4 \over 9}[(F-5D)\mu_u+2(D+F)\mu_s],
\nonumber \\
 \tilde \gamma_{\Sigma^0}^{2(K)} &=& -{2 \over 9}(D+F)(\mu_u-4\mu_s),  \\
 \tilde \gamma_{\Sigma^-}^{2(K)} &=& {8 \over 9}[(2D-F)\mu_u+(D+F)\mu_s],
\nonumber \\
 \tilde \gamma_{\Xi^0}^{2(K)} &=& {2 \over 3}[(D+3F)\mu_u-2(D+F)\mu_s],
\nonumber \\
 \tilde \gamma_{\Xi^-}^{2(K)} &=& -{4 \over 3}[F\mu_u+(D+F)\mu_s], \nonumber \\
 \tilde \gamma_{\Lambda\Sigma^0}^{1(K)}&=&-{4 \over 3\sqrt{3}}(2D+3F)\mu_u,
\nonumber
\end{eqnarray}
for the kaon loops, and by
\begin{eqnarray} \label{tgam2e}
 \tilde \gamma_p^{2(\eta)} &=& 0, \ \ \
 \tilde \gamma_{\Sigma^+}^{2(\eta)} = -{8 \over 9}D(\mu_u-\mu_s), \ \ \ \
 \tilde \gamma_{\Xi^0}^{2(\eta)} = {4 \over 9}(D+3F)(\mu_u-\mu_s),  \nonumber
\\
 \tilde \gamma_n^{2(\eta)} &=& 0 , \ \ \
 \tilde \gamma_{\Sigma^0}^{2(\eta)} = -{2 \over 9}D(\mu_u-4\mu_s), \ \ \
 \tilde \gamma_{\Xi^-}^{2(\eta)} = -{2 \over 9}(D+3F)(\mu_u+2\mu_s), \\
 \tilde \gamma_{\Lambda}^{2(\eta)} &=& 0 , \ \ \
 \tilde \gamma_{\Sigma^-}^{2(\eta)} = {4 \over 9}D(\mu_u+2\mu_s), \ \ \ \ \
 \tilde \gamma_{\Lambda\Sigma^0}^{2(\eta)} = -{1 \over \sqrt{3}}D\mu_u ,
\nonumber
\end{eqnarray}
for the $\eta$ loops.

\section{The expressions for $ L_0$, $\widetilde F$, $ L_1$, and $L_2$}

Let us introduce the following notation

\begin{equation}
 F_0 (m, a) = \left\{
    \begin{array}{ll}
        \sqrt{m^2-a^2} \, [ \pi/2 - {\rm arctan} \,
          (a / \sqrt{m^2-a^2})] & m \geq a \\
&\\
         - \sqrt{a^2-m^2} \, {\rm ln} \,
      [(a +\sqrt{a^2-m^2}\,) / m] & m < a
    \end{array}
 \right.
\end{equation}
where $a$ is an arbitrary parameter. Hereafter, the subscript $(X)$ is
ignored. The function $ L_0(m,\lambda)$ obtained from the Feynman integral for
the graph 3a (or 3e) is then given by
\begin{equation}
L_0(m,\lambda) =
  {\lambda^4 \over (\lambda^2-m^2)^2} \left[ \ {1 \over 3} (\lambda^2+2 m^2)
       + {\lambda m^2 \over \lambda^2-m^2} F_0 (m, \lambda) \right] \, .
\end{equation}
The function $\widetilde F(m,\delta,\lambda)$ obtained when calculating a
Feynman integral for the graph 1b is

\begin{eqnarray} \label{funf}
\pi \widetilde F(m,\delta,\lambda) &=& -{\lambda^4 \over
3(\lambda^2-m^2+\delta^2)^2}
    \left \{ N(m,\delta,\lambda)+{5\lambda^2+2m^2 \over \lambda^2-m^2}\delta +
    {\lambda^2+2m^2 \over (\lambda^2-m^2)^2}\delta^3 \right . \nonumber \\
  &+& {\lambda\delta \over (\lambda^2-m^2)^2(\lambda^2-m^2+\delta^2)}
     \bigg [ 3(2\lambda^2+3m^2)(\lambda^2-m^2)
     -2(\lambda^2-6m^2)\delta^2 \nonumber \\
  &+& \left.\left. {3 m^2 \over \lambda^2-m^2}\delta^4 \right] \,
     F_0 (m, \lambda) \right\} \, ,
\end{eqnarray}
where
\begin{equation}
N(m,\delta,\lambda) = {1 \over (\lambda^2-m^2+\delta^2)}
             \left [ \pi\lambda(\lambda^2+3m^2-3\delta^2)
            - 2(3\lambda^2+m^2-\delta^2) \, F_0 (m, \delta) \right ] \, .
\end{equation}
Similarly, the functions $ L_1(m,\delta,\lambda)$, and
$L_2(m,\delta,\lambda)$ arise from the Feynman integrals for the
graphs 3b (or 3f) and 3c (or 3d), respectively.
We have

\begin{eqnarray} \label{funl}
 L_1(m, \delta,\lambda)&=& {2\lambda^4 \over 3(\lambda^2-m^2+\delta^2)^2}
  \left\{(\lambda^2+2m^2-2\delta^2)-\delta N(m,\delta,\lambda)
- {\lambda^2 \over \lambda^2-m^2}\delta^2 \right. \nonumber \\
  &+& {\lambda \over (\lambda^2-m^2)(\lambda^2-m^2+\delta^2)}
     \bigg [ 3m^2(\lambda^2-m^2)-6\lambda^2\delta^2 \nonumber \\
  &+& \left.\left. {2\lambda^2-3m^2 \over \lambda^2-m^2}\delta^4 \right]
      F_0 (m, \lambda) \right\} \,,
\end{eqnarray}
and

\begin{eqnarray} \label{funlh}
L_2(m, \delta,\lambda) &=&-{2\lambda^4 \over 3(\lambda^2-m^2+\delta^2)^2}
   \left \{ \left[ 2(m^2-\delta^2) \, F_0 (m, \delta)
    -\pi m^3 \right]{1 \over \delta} \right. \nonumber \\
  &+& {\pi\delta \over 2(\lambda+m)}\left[ (\lambda^2+\lambda m+4m^2)
     -{\lambda+2m \over \lambda+m}\delta^2 \right]
     -{\lambda^2(\lambda^2-m^2+\delta^2) \over \lambda^2-m^2} \nonumber \\
  &-& \left . {\lambda \over (\lambda^2-m^2)^2} \left[
    3m^2(\lambda^2-m^2)-(2\lambda^2-3m^2)\delta^2 \right] \
     F_0 (m, \lambda) \right \} \, .
\end{eqnarray}

When $\delta =0$, Eq.\ (\ref{funf}) gives

\begin{equation} \label{fde0}
\widetilde F(m,0,\lambda)= -{\lambda^4 \over 3(\lambda+m)^3} \, ,
\end{equation}
and it follows from Eqs.\ (\ref{funl}) and (\ref{funlh}) that
\begin{eqnarray} \label{lde0}
L_1(m,0,\lambda)  &=& L_2(m,0,\lambda) \, = \,
2L_0(m,\lambda) \nonumber \\
&=& {2\lambda^4\over 3(\lambda^2-m^2)^2}
    \left[ [ \ (\lambda^2+2m^2)+
    {3\lambda m^2 \over \lambda^2-m^2} \, F_0 (m, \lambda)  \ \right] \,.
\end{eqnarray}

\newpage

\begin{figure}
\caption{World sheet picture for the structure of a baryon in the Wilson-loop
approach.}
\label{fig:worldsheet}
\end{figure}

\begin{figure}
\caption{Diagrams with couplings independent of the baryon moments. These
diagrams lead to the non-analytic
$m_s^{1/2}$ corrections to the baryon magnetic moments in the conventional
ChPT. The dashed lines denote the mesons, the single and double solid lines
denote octet and decuplet baryons, respectively. A heavy dot with a meson
line represents a form factor $F(k,v)$ (Eq.(\protect\ref{formfr})), where $k$
is
the meson momentum.}
\label{fig:mesonloop}
\end{figure}

\begin{figure}
\caption{Diagrams with couplings that that depend on the tree-level baryon
moments. These diagrams lead to non-analytic
$m_s \ln m_s$ corrections to the baryon magnetic moments in the
conventional ChPT.}
\label{fig:baryonloop}
\end{figure}

\newpage
\centerline{\hspace*{-1in}
\epsffile{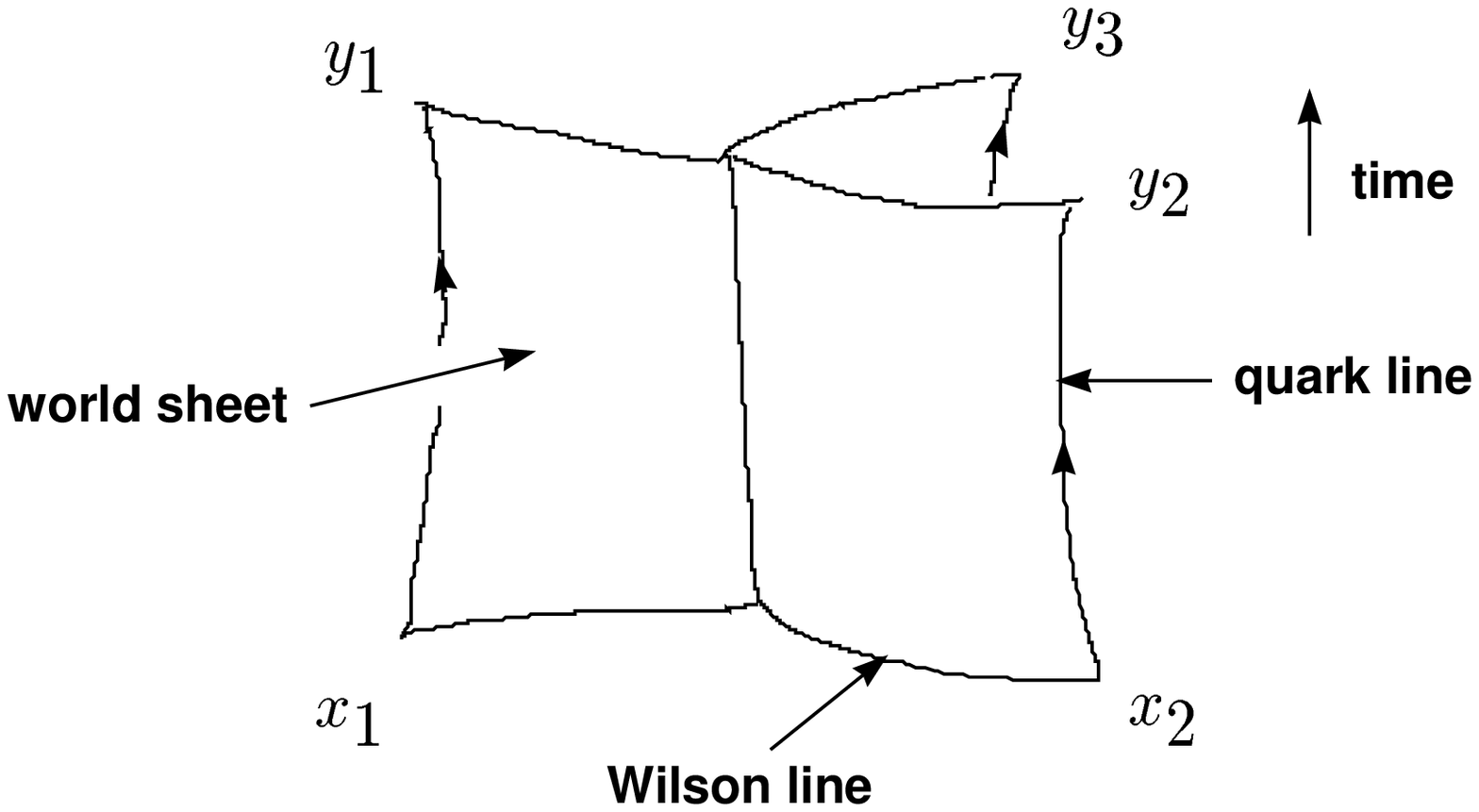}
}

\newpage

\centerline{
\epsffile{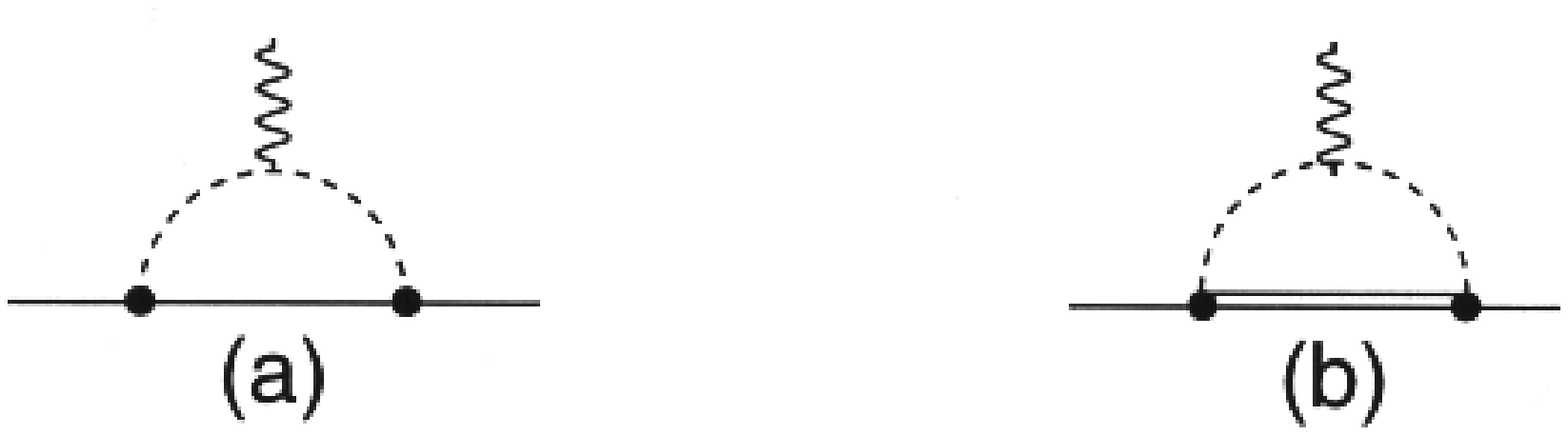}
}

\newpage

\centerline{
\epsffile{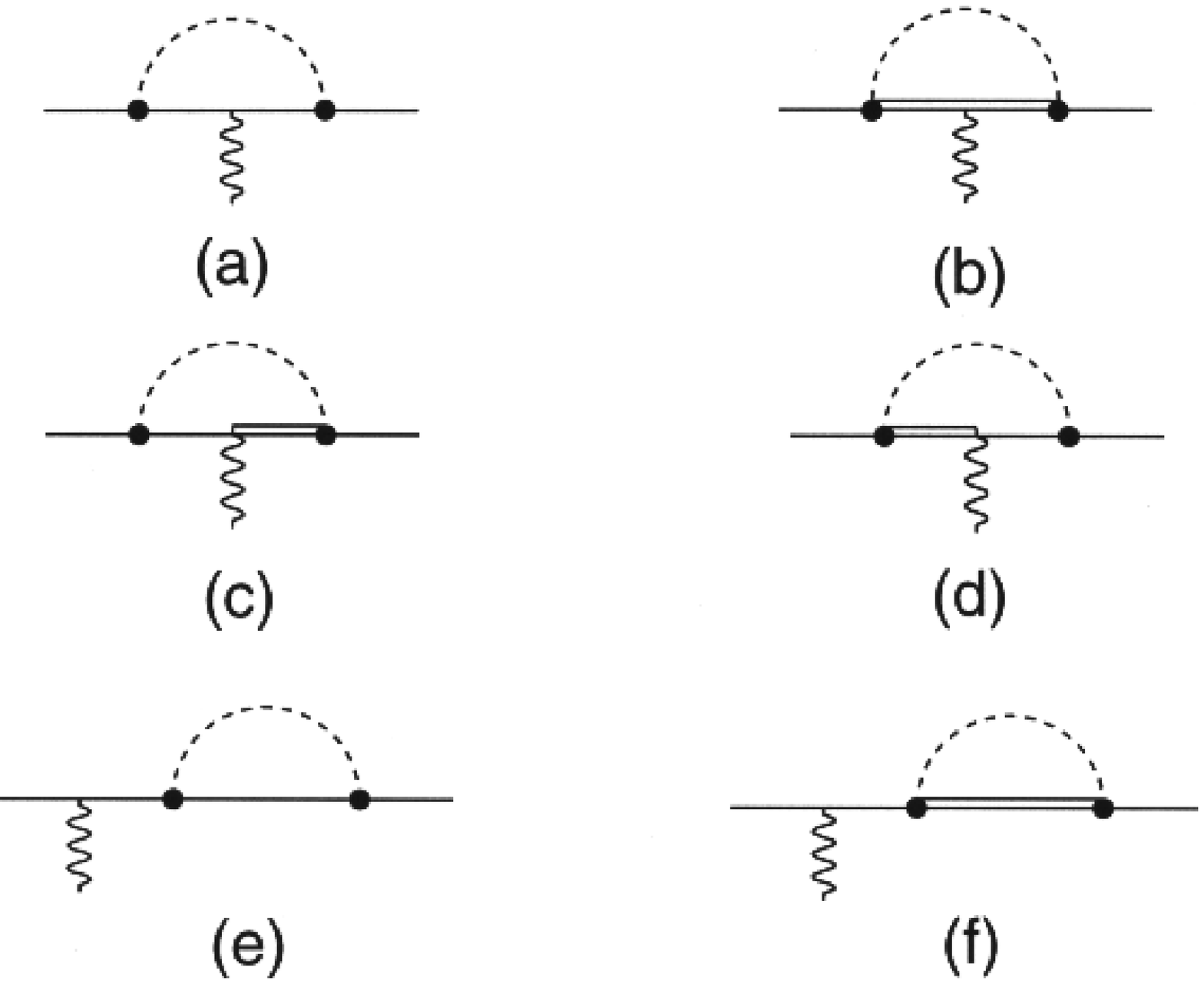}
}

\begin{table}
\caption{ The values in GeV of the matrix elements $\epsilon$ and $\Sigma$
defined in Eqs. (\protect\ref{epsilons}) and (\protect\ref{sigmas})
The matrix elements were evaluated for
$\alpha_s=0.39$, $\sigma=0.18 {\rm GeV}^2$, $m_u=m_d=0.343$ GeV,
and $m_s=0.539$ GeV. }
\label{table:matrix_elements}

\begin{tabular}{crrrrrr}
     Baryon & $\epsilon$ & $\epsilon'$ & $\tilde{\epsilon}$
       & $\Sigma$ & $\Sigma'$ & $\widetilde{\Sigma}$  \\
    \hline
    $N$      & 0.014 & 0.014 & 0.014 & 0.062 & 0.062 &  0.062 \\
    $\Sigma$ & 0.014 & 0.018 & 0.012 & 0.061 & 0.069 &  0.046 \\
    $\Xi$    & 0.016 & 0.012 & 0.018 & 0.053 & 0.046 &  0.068 \\
    $\Omega$ & 0.016 & 0.016 & 0.016 & 0.052 & 0.052 &  0.052
\end{tabular}
\end{table}

\begin{table}
\caption{The magnetic moments from the naive quark model (QM) and the QCD-based
QM with loop corrections. The average deviations from experimental values are
about 0.10 and 0.05 for the QM and the QCD-based QM with loop corrections,
respectively. All moments and deviations are given in units of $\mu_N$.}
\label{table:moments}
\begin{tabular}{crrr}
     $\mu_B$ & QM &  QCD QM\,$+$\,loops & Experiment \\
    \hline
    p                 &  2.728   &  2.720 & $2.793 \pm 0.000$ \\
    n                 & $-$1.818 & $-$1.946 & $-1.913 \pm 0.000$ \\
    $\Sigma^+$        &  2.618   &  2.519 & $2.458 \pm 0.010$ \\
    $\Sigma^-$        & $-$1.019 & $-$1.110 & $-1.160 \pm 0.025$ \\
    $\Sigma^0$        &  0.800   &  0.705 &  $-$ \\
    $\Lambda$         & $-$0.580 & $-$0.608 & $-0.613 \pm 0.004$ \\
    $\Xi^0$           & $-$1.380 & $-$1.316 & $-1.250 \pm 0.014$ \\
    $\Xi^-$           & $-$0.470 & $-$0.582 & $-0.651 \pm 0.003$ \\
    $\Sigma^0\Lambda$ &  1.575   &  1.559  &  $\pm1.610 \pm 0.08$\\
       \hline
    $\mu_u$ & 1.818 & 2.083 & $-$ \\
    $\mu_s$ & $-$0.580 & $-$0.656 & $-$ \\
    $\lambda$ (in MeV) & $-$ & 407 & $-$ \\
  \end{tabular}
\end{table}

\begin{table}
\caption{Detailed breakdown of the contributions of the loop integrals
to the fitted magnetic moments of the octet baryons (in $\mu_N$). Those
contributions are evaluated at $F=0.5$, $D=0.75$, $ {\cal C} =-1.5$. As shown
in Table I, a best fit is obtained at $\mu_u=2.083$, $\mu_s=-0.656$ and the
natural cutoff $\mu =407$ MeV. $\Delta \mu_B$ stands for a deviation from
the experimently measured value.}
\label{table:loop_details}
\begin{tabular}{lrrrrrrrrr}
     $\mu_B$ & $\mu_u$, $\mu_s$ & ${\tilde\Delta} \mu_i$ & $m_s^{1/2(N)}$ & ln
$m_s^{(N)}$ & $m_s^{1/2 (\Delta)}$ & ln $m_s^{(\Delta)}$ & Loops & $\mu_B$ & $
\Delta \mu_B $ \\
    \hline
    p       &  3.124 & $-$0.500 & 0.416  & $-$0.729  &  0.057 & 0.353 & 0.096 &
2.720 & $-$0.072 \\
    n       & $-$2.083 & 0.427 & $-$0.381 & 0.409 & $-$0.067 & $-$0.253 &
$-$0.291 & $-$1.946 &  $-$0.033 \\
  $\Sigma^+$  &  2.995 & $-$0.428 & 0.273  & $-$0.536 & $-$0.002 & 0.217 &
$-$0.048 & 2.519 & 0.061 \\
    $\Sigma^-$  & $-$1.170 & 0.139 &$-$0.217 & 0.188 & 0.021 & $-$0.072 &
$-$0.080 & $-$1.110 &  0.050 \\
    $\Sigma^0$  & 0.913 & $-$0.145 & 0.028  & $-$0.174 &0.010 & 0.073 &
$-$0.064 & 0.705  &  \\
    $\Lambda$   & $-$0.656 & 0.085 &$-$0.028 & 0.073 & $-$0.010 & $-$0.073 &
$-$0.037 & $-$0.608 & 0.005 \\
    $\Xi^0$     & $-$1.569 & 0.271 &$-$0.043 & 0.128 & $-$0.043 & $-$0.059 &
$-$0.017 & $-$1.316 & $-$0.066 \\
  $\Xi^-$    & $-$0.528 & $-$0.056 & $-$0.048 & 0.061 & 0.033 & $-$0.044 &
0.002 & $-$0.582 &  0.069 \\
    $\Sigma^0\Lambda$ & 1.804  & $-$0.378 & 0.229 & $-$0.268 & 0.058 & 0.114 &
0.134 & 1.559 & $-$0.051 \\
  \end{tabular}

\end{table}

\end{document}